\documentclass[11pt]{article}

\begin{document}
\hoffset=-0.25in
\voffset=-0.75in

\title{\bf Conformal Gravity Challenges String Theory}

\author{Philip D. Mannheim \\Department of Physics,
University of Connecticut, Storrs, CT 06269, USA \\
e-mail: philip.mannheim@uconn.edu\\ \\
July 16, 2007}

\date{}
\maketitle

\noindent
{\bf Abstract.} The cosmological constant problem and the compatibility of gravity with quantum mechanics are the two most pressing problems in all of gravitational theory. While string theory nicely addresses the latter,  it has so far failed to provide any compelling solution to the former. On the other hand, while conformal gravity nicely addresses the cosmological constant problem (by naturally quenching the amount by which the cosmological constant gravitates rather than by quenching the cosmological constant itself), the fourth order derivative conformal theory has long been thought to possess a ghost when quantized. However, it has recently been shown by Bender and Mannheim that not only do theories based on fourth order derivative equations of motion not have ghosts, they actually never had any to begin with, with the apparent presence of ghosts being due entirely to treating operators which were not Hermitian on the real axis as though they were. When this is taken care of via an underlying $\cal P$$\cal T$ symmetry that such theories are found to possess, there are then no ghosts at all and the $S$-matrix is fully unitary.  Conformal gravity is thus advanced as a fully consistent four-dimensional alternative to ten-dimensional string theory.



\section{The Motivation for Einstein Gravity}

In this paper we compare and contrast conformal gravity with string theory, and to do so we need to go back to the very beginnings of general relativity. As we recall, following his development of special relativity, Einstein was faced with the problem of making gravity (as then expressed through Newton's Law of Gravity) compatible with the relativity principle. Einstein achieved this by generalizing the gravitational potential $\phi$ to the metric $g_{\mu\nu}$, by requiring that the metric  couple to matter covariantly so that test particles would then move on the geodesics associated with the metric, and by specifying a particular set of equations, viz. the Einstein equations 
\begin{equation}
-\frac{1}{8\pi G}\left(R_{\mu\nu} -\frac{1}{2}g_{\mu\nu}R^{\alpha}_{\phantom{\alpha}\alpha}\right)=T_{\mu\nu}
\label{1}
\end{equation}
for a source with energy-momentum tensor $T_{\mu\nu}$,  which were to fix the Christoffel symbols needed for the geodesics. As constructed Einstein's general relativity theory consists of two separate ingredients, one kinematic and the other dynamic, the kinematic one being that gravity is a covariant metric theory, and the dynamic one being that the equation of motion for the metric is to be given by Eq. (\ref{1}). The kinematic ingredient is very general as it embodies the equivalence principle, and would hold for any covariant pure metric theory of gravity, but the dynamic ingredient was motivated solely by the purely phenomenological requirement that the theory reduce to the second order Poisson equation $\nabla^2\phi =4\pi G\rho$ in the weak gravity limit. With the Einstein equations not only recovering Newton's Law of Gravity but also leading to general relativistic corrections to it  which were then spectacularly confirmed (the three classic solar system tests), the general view of the community is that the issue of what is the correct theory of gravity is now settled. 

\section {The Shortcomings of Einstein Gravity}

Despite these successes there are some very serious concerns with the above analysis. Specifically we need to ask what is so special about the second order Poisson equation (i.e. where did it come from) since the use of it was motivated purely by phenomenological reasoning and not via any fundamental principle. And secondly, if it is to be phenomenological viability which is to be the criterion, we need to ask whether the Einstein theory is in fact the only theory which then meets the three classic tests. And as we shall see in our discussion of conformal gravity below, it is actually possible to bypass both the second order Poisson equation and the Einstein equations altogether and yet still recover Newton's Law of Gravity and its covariant Schwarzschild solution generalization, with the Einstein theory only being sufficient to give Newton and Schwarzschild but not necessary. 

Beyond this, we also note that when Einstein gravity is extended beyond its solar system origins, no matter in which way it is extended additional concerns arise. When Einstein gravity is extended to galactic distance scales we get the dark matter problem. When Einstein gravity is extended to cosmological distance scales we get the cosmological constant or dark energy problem. When Einstein gravity is extended to strong gravitational fields we get the singularity problem. And finally, when Einstein gravity is extended far off  the mass shell we get the renormalization problem. For none of these problems is there as yet any solution which has been experimentally validated. The case for dark matter is made solely by assuming the a priori validity of Einstein gravity and then arguing that its failure to fit astrophysical data with luminous sources alone is evidence for the existence of dark matter, dark matter which has yet to be directly detected in any of the extensive dark matter searches which have been going on for many years now. The cosmological constant problem is even more severe since the needed value as inferred from the application of standard gravity to cosmology is 60 to 120 orders of magnitude less than the value suggested by microscopic elementary particle physics. With regard to singularities, not only are there no data which provide direct evidence for the existence in nature of event horizons or trapped surfaces (or even whether the mass concentrations in galactic centers have radii less than their Schwarzschild radii), it is not clear whether the existence of singularities in the fabric of spacetime is a property of nature or an indication of the breakdown of the theory. Finally, to resolve the renormalizability problem it has been found necessary to generalize the theory to a superstring theory  which introduces two further ingredients for which there is also no experimental evidence, namely the existence of ten spacetime dimensions and the existence of a supersymmetry which gives all known particles as yet undetected superpartners.

As introduced, string theory only exacerbates the cosmological constant problem since it is very hard to secure the needed miniscule value for the cosmological constant $\Lambda$ in a theory which has to make the superpartners of the ordinary particles so heavy as to account for their non-detection in accelerator experiments. In fact, it is very hard to secure so miniscule a value for $\Lambda$ at all, and appeal is now being made to an anthropic origin for such a value for $\Lambda$. As such, the use of the anthropic principle to fix the values of physical parameters such as $\Lambda$ requires that three conditions hold: firstly that we have reached a brick wall in trying to explain the value of $\Lambda$ by conventional means, secondly that there is a theory which admits of an entire range of allowed values for $\Lambda$, and thirdly that in such a theory there is only a very narrow range of values for $\Lambda$ for which intelligent observers could emerge. As regards the first condition, it is almost impossible to demonstrate that we have indeed reached a brick wall or that we have reached a limit to knowledge. Rather all we can say is that we have run out of ideas. As regards the second condition, with the development of the string landscape with its at least $10^{500}$ solutions, string theory does at least provide some range of allowed values for $\Lambda$ [with the landscape converting a manifest vice (viz. too many solutions) into a possible virtue]. However, whether any single one of these $10^{500}$ solutions looks anything like the universe we observe remains to be seen, so it is not at all clear if the third condition is actually met anywhere in the landscape, or if there is any region in the landscape where $\Lambda$ is very small and the supersymmetry  superpartners are very heavy. Nonetheless, no matter what mechanism might make $\Lambda$ so small, as we will show below there is actually a direct  observational test for such a low value. In fact we will see that this same test will also test conformal gravity, a theory that resolves the brick wall problem by asserting that the brick wall arises solely because of the use of Einstein gravity, and that no amount of effort will ever pierce the brick wall  and solve the cosmological constant problem as long as one continues to use the standard theory.

\section {The Non-Uniqueness of Einstein Gravity}

Currently the literature abounds with generalizations of Einstein gravity, almost all of which have been developed to address the dark matter and/or dark energy problems. Essentially all of these generalizations  are based on the ``Einstein Plus" notion that one can add additional terms to Eq. (\ref{1}) provided they are all negligible in the weak gravity solar system limit. Such generalizations include the addition of new functions of the metric [$f(R)$ theories] or new fields such as scalars or vectors. It is not clear why there should be any interest in such theories since none of them shed light on the cosmological constant problem, and unless any of them is shown to descend from string theory, none of them would lead to a consistent quantum gravity. 

Beyond the above type of non-uniqueness, there is an additional, quite different and altogether more serious type of non-uniqueness in the Einstein theory, one related to the status of the second order Poisson equation. Specifically, even though $\phi=-\beta/r$ is a solution to $\nabla^2\phi=\rho$, the Newtonian potential is not uniquely tied to the second order Poisson equation since $\phi=-\beta/r+\gamma r$ is a solution to the fourth order $\nabla^4\phi=\rho$. Moreover, even while the $\phi=-\beta/r+\gamma r$ solution reduces to $\phi=-\beta/r$ for small enough $r$, there is no limit in which $\nabla^4\phi=\rho$ reduces to $\nabla^2\phi=\rho$. Consequently Newton's Law of Gravity can be recovered in theories in which the second order Poisson equation does  not appear at all. Thus one can bypass the second order Poisson equation altogether and thus remove the very first step in a chain of reasoning which ultimately takes us to ten-dimensional string theory and the cosmological constant problem.

Moreover, precisely the same analysis can be applied to the three classic relativistic correction tests to Newton. Specifically,  with all of these tests being embodied in the $R_{\mu\nu}=0$ exterior Schwarzschild  solution to Eq. (\ref{1}), all of these tests can equally be met in theories in which instead some derivative function of the Ricci tensor vanishes since its vanishing can be achieved by the vanishing of the Ricci tensor itself. And indeed this is precisely what occurs in the fourth order derivative conformal gravity theory, since the Einstein-Hilbert action $I_{\rm EH}=-(1/16 \pi G)\int d^4x (-g)^{1/2}R^{\alpha}_{\phantom {\alpha}\alpha}$ is replaced by the Weyl action $I_W=-\alpha_g\int d^4x (-g)^{1/2}C_{\lambda\mu\nu\kappa} C^{\lambda\mu\nu\kappa}$ where $C_{\lambda\mu\nu\kappa}$ is the Weyl tensor and the coupling constant $\alpha_g$ is dimensionless, with Eq. (\ref{1}) being replaced by 
\begin{equation}
4\alpha_g W^{\mu\nu}=T^{\mu\nu}~~
\label{2}
\end{equation}
where $W^{\mu\nu}$ is given by
\begin{eqnarray}
W^{\mu \nu}&= &
\frac{1}{2}g^{\mu\nu}(R^{\alpha}_{\phantom{\alpha}\alpha})   
^{;\beta}_{\phantom{;\beta};\beta}+
R^{\mu\nu;\beta}_{\phantom{\mu\nu;\beta};\beta}                     
 -R^{\mu\beta;\nu}_{\phantom{\mu\beta;\nu};\beta}                        
-R^{\nu \beta;\mu}_{\phantom{\nu \beta;\mu};\beta}                          
 - 2R^{\mu\beta}R^{\nu}_{\phantom{\nu}\beta}                                    
+\frac{1}{2}g^{\mu\nu}R_{\alpha\beta}R^{\alpha\beta}
\nonumber \\
&&-\frac{2}{3}g^{\mu\nu}(R^{\alpha}_{\phantom{\alpha}\alpha})          
^{;\beta}_{\phantom{;\beta};\beta}                                              
+\frac{2}{3}(R^{\alpha}_{\phantom{\alpha}\alpha})^{;\mu;\nu}                           
+\frac{2}{3} R^{\alpha}_{\phantom{\alpha}\alpha}
R^{\mu\nu}                              
-\frac{1}{6}g^{\mu\nu}(R^{\alpha}_{\phantom{\alpha}\alpha})^2.
\label{3}
\end{eqnarray}                                 
As can be seen, $R_{\mu\nu}=0$ is an exact exterior solution to Eq. (\ref{2}). Thus just like the Newtonian potential, the Schwarzschild solution is not unambiguously tied to the standard Newton/Einstein gravitational theory. With Eq. (2) being found to reduce to a fourth order Poisson equation in the weak gravity limit and with Eq. (\ref{2}) never reducing to Eq. (\ref{1}) in any limit, we see that in the conformal theory both the second order Poisson equation and its standard second order Einstein equation generalization are bypassed completely, with conformal gravity thus not being in the ``Einstein Plus" category at all. Rather, it represents a novel and altogether different approach to gravity. With the fourth order Poisson equation leading to modifications to the Newton potential at large distances, modifications which grow rather than fall with distance, conformal gravity modifies standard gravity in precisely the region where the standard theory encounters the need for dark matter, with conformal gravity having been found fully capable of giving an accounting of galactic rotation curve systematics without any need to invoke dark matter at all \cite{Mannheim2006}. It is thus the change in the extrapolation of solar system wisdom to galaxies which solves the dark matter problem.

\section{Conformal Cosmology}

However, it is in its application to cosmology that conformal gravity really comes into its own, since it is the very existence of its underlying conformal invariance [viz. invariance of the geometry under any and all local conformal transformations of the form $g_{\mu\nu}\rightarrow e^{2\alpha(x)}g_{\mu\nu}$] which forces the gravitational action to be uniquely given by the Weyl action $I_W$, with any fundamental cosmological constant then expressly being forbidden.\footnote{With the underlying conformal invariance leading uniquely to Eq. (\ref{2}) and its weak gravity fourth order Poisson equation limit, conformal gravity also shows that there is in fact nothing special about $\nabla^2\phi=\rho$ at all. Rather, it is $\nabla^4\phi=\rho$  which is special since there is a specific symmetry, viz. conformal invariance, which uniquely picks it out.} The conformal symmetry thus forces the cosmological constant to be zero at the level of the action. 

This same conformal symmetry also forces the Planck mass to be zero as well [no $(1/16 \pi G) R^{\alpha}_{\phantom{\alpha}\alpha}$ term in the Lagrangian], to thus require gravitational mass scales to be induced dynamically, and thereby allow global cosmology to be controlled by an induced scale which could be completely different from the induced Cavendish one which is to control local physics. In addition this same  conformal symmetry also sharply restricts the form of the matter action, with the prototypical conformal invariant action which consists of a massless fermion $\psi(x)$ and a spontaneous symmetry breaking scalar field $S(x)$ being of the form 
\begin{equation}
I_M=-\int d^4x(-g)^{1/2}\left[\frac{1}{2}S^{;\mu}
S_{;\mu}-\frac{1}{12}S^2R^\mu_{\phantom         
{\mu}\mu}
+\lambda S^4
+i\bar{\psi}\gamma^{\mu}(x)[\partial_\mu+\Gamma_\mu(x)]             
\psi -hS\bar{\psi}\psi\right].
\label{4}
\end{equation}                                 
In the presence of a perfect fluid of fermions and a constant expectation value $S_0$ for the symmetry breaking scalar field, the energy-momentum tensor associated with the matter $I_M$ is found to be of the form
\begin{equation}
T_{\mu\nu}=(\rho+p)U_{\mu}U_{\nu}+pg_{\mu\nu}-\frac{1}{6}S_0^2\left(R_{\mu\nu}
-\frac{1}{2}g_{\mu\nu}R^\alpha_{\phantom{\alpha}\alpha}\right)-g_{\mu\nu}\lambda S_0^4,
\label{5}
\end{equation}                                 
 a form that thus contains a dynamically induced cosmological constant $\Lambda=\lambda S_0^4$.  Then, since the Weyl tensor vanishes identically in geometries such as the homogeneous and isotropic Robertson-Walker geometry, in conformal cosmology the left-hand side of Eq. (\ref{2}) vanishes identically, with the cosmology thus being described by the equation of motion $T_{\mu\nu}=0$, viz. by 
\begin{equation}
\frac{1}{6}S_0^2\left(R_{\mu\nu}
-\frac{1}{2}g_{\mu\nu}R^\alpha_{\phantom{\alpha}\alpha}\right)=(\rho+p)U_{\mu}U_{\nu}+pg_{\mu\nu}-g_{\mu\nu}\lambda S_0^4.
\label{6}
\end{equation}                                 
As such, conformal cosmology thus looks precisely like a standard cosmology with a perfect matter fluid and a cosmological constant save only that Newton's constant has been replaced by an effective Newton constant of the form 
\begin{equation}
G_{\rm eff}=-\frac{3}{4\pi S_0^2}.
\label{7}
\end{equation}                                 
Since the induced $G_{\rm eff}$ is negative rather than positive, in conformal cosmology gravity is repulsive rather than attractive, with the cosmology thus automatically being an accelerating rather than a decelerating one in all epochs, to naturally explain the accelerating universe without fine-tuning. Additionally, since $G_{\rm eff}$ is negative, there is no initial singularity (i.e. the cosmology expands from a large but finite initial temperature), with gravity thus being able to protect itself from its own singularities.

However, of even more interest is the fact that the induced $G_{\rm eff}$ behaves as $-1/S_0^2$, to thus become smaller the larger the symmetry breaking scale $S_0$ might be. The larger the induced cosmological constant then, the less it gravitates, with conformal gravity providing a natural quenching not present in the standard theory. Moreover, one can explicitly show \cite{Mannheim2006} that no matter what the numerical values of the scales associated with the symmetry breaking, the structure of Eq. (\ref{6}) is such that the quantity $\bar{\Omega}_{\rm \Lambda} =8\pi G_{\rm eff}\Lambda/3H^2$ which describes how the cosmological constant couples to the geometry is constrained to lie in the range $0 \leq \bar{\Omega}_{\rm \Lambda} \leq 1$ in all epochs. Thus no matter how large the $\Lambda$ induced by symmetry breaking might be, its coupling to cosmology is completely under control. Finally, through use of Eq. (\ref{6}), fits to the accelerating universe supernovae data have been obtained \cite{Mannheim2006} which are every bit as high in quality as those obtained in an $\Omega_M=0.3$, $\Omega_{\Lambda}=0.7$ standard cosmology. Moreover, unlike the situation in the standard model, because of the $G_{\rm eff}$ quenching, in the conformal case the fits are not fine-tuned and can accommodate as large a cosmological constant term as one wants. With the conformal cosmology being accelerating in all epochs, and with the standard model having been fine-tuned to only be accelerating in the current (redshift $z$ less than one) epoch, conformal gravity predicts continuing acceleration above $z=1$ while the standard theory predicts a switch over to deceleration above $z=1$. Exploration of the $z>1$ Hubble plot will not only provide a clear cut way of discriminating between standard cosmology and its conformal challenger, since the use of anthropic argumentation in the landscape picture is supposed to also lead to $\Omega_{\Lambda}=0.7$, it too requires deceleration above $z=1$ and is thus amenable to falsification.

\section{Quantum Conformal Gravity}

With the Weyl action $I_W$ possessing a dimensionless gravitational coupling constant $\alpha_g$, unlike the standard Einstein-Hilbert action with its dimension two $1/16 \pi G$ factor, conformal gravity is immediately power counting renormalizable, with its behavior far off the mass shell thus being completely under control. However, with the wave equation for fluctuations around flat spacetime obeying \cite{Mannheim2006} the fourth order wave equation $(-\partial_t^2+\nabla^2)^2K(\bar{x},t)=0$ for each independent component $K_{\mu\nu}=h_{\mu\nu}-(1/4)\eta_{\mu\nu}h^{\alpha}_{\phantom{\alpha}\alpha}$ of the perturbed metric, the theory is thought to have a ghost state and thus not be unitary. To see the nature of this supposed disease as clearly as possible, it is instructive to look at the second plus fourth order wave equation $(-\partial_t^2+\nabla^2)(-\partial_t^2+\nabla^2-M^2)K(\bar{x},t)=0$, since its momentum space propagator takes the form
\begin{equation}
D(k^2,M)=
\frac{1}{k^2(k^2+M^2)}=\frac{1}{M^2}\bigg{[}\frac{1}{k^2} -\frac{1}{(k^2+M^2)}\bigg{]},
\label{8}
\end{equation}                                 
with the relative minus sign between the two terms suggesting the presence of a negative residue ghost state -- i.e. the same mechanism which makes the conformal theory far better behaved in the ultraviolet than the standard theory appears to do so by having the regular graviton be accompanied by a ghost graviton. However, before one can assert that the structure of Eq. (\ref{8}) implies the presence of a ghost, one must first construct the correct Hilbert space for the theory and identify the appropriate inner product. This procedure has been carried through by Bender and Mannheim \cite{Bender2007} in a prototypical case with a quite surprising result being obtained, namely that  the minus sign in Eq. (\ref{8}) is not symptomatic of a ghost at all but can actually appear in theories with strictly positive norm. 

The approach of \cite{Bender2007} was to descend to quantum mechanics by setting $K(\bar{x},t)=z(t) e^{i\bar{k}\cdot\bar{x}}$, with the wave equation reducing to the prototypical Pais-Uhlenbeck (PU) fourth order oscillator equation $d^4z/dt^4+(\omega_1^2+\omega_2^2)d^2z/dt^2+\omega_1^2\omega_2^2z=0$ where we have set $\omega_1^2+\omega_2^2=2k^2+M^2$, $\omega_1^2\omega_2^2=k^4+k^2M^2$. With this wave equation being obtainable from the action $I_{\rm PU}=(\gamma/2)\int dt[ \ddot{z}^2-(\omega_1^2+\omega_2^2)\dot{z}^2+\omega_1^2\omega_2^2z^2]$, from this $I_{\rm PU}$ Mannheim and Davidson \cite{Mannheim2000} used the method of Dirac constraints to construct the associated Hamiltonian
\begin{equation}
H=\frac{p_x^2}{2\gamma}+p_zx+\frac{\gamma}{2}\left(\omega_1^2+\omega_2^2
\right)x^2-\frac{\gamma}{2}\omega_1^2\omega_2^2z^2
\label{9}
\end{equation}
where $x$ has replaced $\dot{z}$, and where $p_x$ and $p_z$ are the conjugates of $x$ and $z$ according to $[x,p_x]=i$, $[z,p_z]=i$, with the theory thus having a two-oscillator structure. 

With the substitutions
\begin{eqnarray}
z&=&a_1+a_1^\dagger+a_2+a_2^\dagger,\qquad 
p_z=i\gamma\omega_1\omega_2^2
(a_1-a_1^\dagger)+i\gamma\omega_1^2\omega_2(a_2-a_2^\dagger),
\nonumber\\
x&=&-i\omega_1(a_1-a_1^\dagger)-i\omega_2(a_2-a_2^\dagger),\qquad
p_x=-\gamma\omega_1^2 (a_1+a_1^\dagger)-\gamma\omega_2^2(a_2+a_2^\dagger)
\label{10}
\end{eqnarray}
yielding a Hamiltonian and Fock-space commutators of the form
\begin{equation}
H=2\gamma(\omega_1^2-\omega_2^2)(\omega_1^2 a_1^\dag a_1-\omega_2^2a_2^\dag a_2)
+\frac{1}{2}(\omega_1+\omega_2),
\label{11}
\end{equation}
\begin{equation}
[a_1,a_1^\dag]=\frac{1}{2\gamma\omega_1
\left(\omega_1^2-\omega_2^2\right)}, \qquad
[a_2,a_2^\dag]=-\frac{1}{2\gamma\omega_2
\left(\omega_1^2-\omega_2^2\right)},
\label{12}
\end{equation}
we see that if we take  $a_1$ and $a_2$ to annihilate the no-particle state $|\Omega\rangle$ according to $a_1|\Omega\rangle=0$, $a_2|\Omega\rangle=0$, we find that the state $|\Omega\rangle$ is the ground state of the system with energy $E_0=(1/2)(\omega_1+\omega_2)$, but on taking $\gamma>0$ and $\omega_1^2>\omega_2^2$ (for definitiveness), we find that the excited state $a_2^\dag|\Omega\rangle$ which lies at energy $\omega_2$ above the ground state has a Dirac norm $\langle\Omega |a_2a_2^\dag|\Omega\rangle$ which is negative. This then is the ghost problem of higher derivative theories, a problem from which there would appear to be no escape.

However, it turns out \cite{Bender2007} that there is a hidden flaw in the above reasoning. To see what the flaw is, it is instructive to make a standard wave mechanics representation $p_z=-i\partial/\partial z$, $p_x=-i\partial/\partial x$, and set up a Schr$\ddot{\rm o}$dinger equation $H\psi_n(z,x)=E_n\psi_n(z,x)$, to find that the state with energy $E_0=(1/2)(\omega_1+\omega_2)$ has eigenfunction
\begin{equation}
\psi_0(z,x)=\exp\bigg[\frac{\gamma}{2}(\omega_1+\omega_2)
\omega_1\omega_2z^2
+i\gamma\omega_1\omega_2zx -\frac{\gamma}{2}(\omega_1+\omega_2)x^2
\bigg],
\label{13}
\end{equation}
while the states with higher energy have eigenfunctions which are polynomial functions of $z$ and $x$ times this $\psi_0(z,x)$. As constructed, both $\psi_0(z,x)$ and all the higher energy eigenfunctions are not normalizable on the real $z$ axis. Consequently, the coordinate space realization of $p_z$ as $-i\partial/\partial z$ cannot be Hermitian on the real axis. Since one can only use the realization $p_z=-i\partial/\partial z$ when the $[z,p_z] $ commutator acts on well-behaved test functions, we see that such well-behaved functions cannot be taken to lie on the real $z$ axis. Rather,  we must take $z$ to lie on the imaginary axis, and on so doing  find that the eigenfunctions are then normalizable. Since the operators $z$ and $p_z$ are not Hermitian on the real axis (rather they are anti-Hermitian there), the Fock space realizations for them in Eq. (\ref{10}) are not valid, with Eqs. (\ref{11}) and (\ref{12}) simply not correctly characterizing the theory.

To determine what the appropriate Fock space is for the theory, it is more convenient to work with real coordinates. We thus set $y=-iz$ and  $q=ip_z$, with the Hamiltonian of Eq. (\ref{9}) then being found to take the form 
\begin{equation}
H=\frac{p^2}{2\gamma}-iqx+\frac{\gamma}{2}\left(\omega_1^2+\omega_2^2
\right)x^2+\frac{\gamma}{2}\omega_1^2\omega_2^2y^2,
\label{14}
\end{equation}
where for notational simplicity we have replaced $p_x$ by $p$. As such, the operators $p$, $x$, $q$ and $y$ all act on the real $x$ and $y$ axes and are all Hermitian in the conventional Dirac sense and obey the commutator algebra $[x,p]=i$, $[y,q]=i$. However, because of its $-iqx$ term the Hamiltonian is manifestly not Hermitian, and thus despite its appearance, the original Hamiltonian of Eq. (\ref{9}) could not have been Hermitian either. The ghost thus only appeared in the theory because one treated the Hamiltonian as though it were (c.f. Eq. (\ref{11})).

Despite its not being Hermitian, the Hamiltonian of Eq. (\ref{14}) still has an eigenspectrum which is strictly real since it falls into the class of $\cal P$$\cal T$ symmetric theories considered by Bender and collaborators (see e.g. \cite{Bender2007b}). To see the  $\cal P$$\cal T$ symmetry we assign the  $p$, $x$, $q$ and $y$ dynamical operators the respective $\cal P$ eigenvalues  $(-,-,+,+)$ and the respective $\cal T$ eigenvalues  $(-,+,+,-)$; and with these operators thus having the respective $\cal P$$\cal T$ eigenvalues  $(+,-,+,-)$, the Hamiltonian of Eq. (\ref{14}) is thus seen to obey $[{\cal P \cal T}, H]=0$. 

In order to construct the Hilbert space which is to be associated with the Hamiltonian of Eq. (\ref{14}), it is convenient \cite{Bender2007b} to construct the so-called $\cal C$ operator which is to obey the relations  
${\cal C}^2=1$, $[{\cal C}, {\cal P \cal T}]=0$, $[{\cal C},H]=0$. On introducing the Hermitian operator ${\cal Q}=\alpha pq+\beta xy$, we find \cite{Bender2007} that the operator ${\cal C}$ will be given as ${\cal C}=e^{-{\cal Q}}{\cal P}$ provided the parameters $\alpha$ and $\beta $ are given by $\beta=\gamma^2\omega_1^2\omega_2^2\alpha$,  $\sinh(\sqrt{\alpha\beta})=2\omega_1\omega_2/(\omega_1^2-\omega_2^2)$.

As is characteristic of $\cal P$$\cal T$ symmetric theories, the non-unitary similarity transformation with $S=e^{-{\cal Q}/2}$ will bring the Hamiltonian to a form in which it is  manifestly Hermitian:
\begin{equation}
{\tilde H}=e^{-{\cal Q}/2}He^{{\cal Q}/2}
=\frac{p^2}{2\gamma}+\frac{q^2}{2\gamma\omega_1^2}+
\frac{\gamma}{2}\omega_1^2x^2+\frac{\gamma}{2}\omega_1^2\omega_2^2y^2,
\label{15}
\end{equation}
with the theory now taking the form of two completely normal harmonic oscillators, each with a bounded, positive Dirac-normed spectrum of energy eigenstates which obey the completely conventional $\langle \tilde{n}|\tilde{m}\rangle=\delta_{\tilde{m},\tilde{n}}$, $\sum_{\tilde{n}}|\tilde{n}\rangle\langle \tilde{n}|=1$,
with the theory thus being manifestly unitary. With the mapping of Eq. (\ref{15}) being with $e^{-{\cal Q}/2}$, on recalling that ${\cal Q}$ is Hermitian, we see that the eigenstates of the Hamiltonian of Eq. (\ref{14}) must thus obey 
\begin{equation}
\langle n |e^{-{\cal Q}}| m\rangle=\delta_{m,n}, \qquad \sum_{n}|n\rangle\langle n|e^{-{\cal Q}}=1,
\label{16}
\end{equation}
with it being $\langle n |e^{-{\cal Q}}$ which is thus the conjugate of  $|n\rangle$ and not  $\langle n |$ itself, with the underlying ${\cal P}$${\cal T}$ symmetry thus fixing the appropriate inner product for the theory.

For this theory we can construct wave functions of the form $\psi_0(x,y)=\langle x,y|0\rangle= \exp\left[-(\gamma/2)(\omega_1+\omega_2)\omega_1\omega_2y^2
-\gamma\omega_1\omega_2yx -(\gamma/2)(\omega_1+\omega_2)x^2\right]$. However, since these wave functions are real, their conjugates cannot be given by complex conjugation. For the conjugates, we need to solve the Schr$\ddot{\rm o}$dinger equation $\psi_n(x,y)H=\psi_n(x,y)E_n$ where $H$ acts to the left. For this case the correct representation of $q$ is $q=+i\partial/\partial y$, to give the wave function $\psi^c_0(x,y)= \exp\left[-(\gamma/2)(\omega_1+\omega_2)\omega_1\omega_2y^2
+\gamma\omega_1\omega_2yx -(\gamma/2)(\omega_1+\omega_2)x^2\right]$. Comparing with $\psi_0(x,y)$, we  see that the conjugates are given not by $\langle n|x,y\rangle$, but by $\psi^c_n(x,y)=\langle n|-x,y\rangle=\langle n|{\cal P}|x,y\rangle$ where ${\cal P}$ is the parity operator. For the propagator, we can insert a complete set of states by writing  $H=\sum_{n}|n\rangle E_n\langle n|e^{-{\cal Q}}=\sum_{n}|n\rangle E_n\langle n|{\cal C}{\cal P}$. However, since ${\cal C}$ commutes with $H$ and obeys ${\cal C}^2=1$, the states $\langle n|$ are eigenstates of ${\cal C}$ with eigenvalues ${\cal C}_n$ which are either plus or minus one. The propagator is thus given by 
\begin{equation}
\langle x^{\prime}, y^{\prime}|e^{-iHt}|x,y \rangle=
\sum_n \psi_n(x^{\prime},y^{\prime})e^{-iE_n t}{\cal C}_n\psi_n^c(x,y),
\label{17}
\end{equation}
with it thus being the appropriate ${\cal C}_n$ eigenvalue which generates the relative minus sign in Eq. (\ref{8}) even though the associated Hilbert space states all have positive norm. Similarly, the orthonormality relations are given by $\langle n |e^{-Q}|m \rangle=\int dxdy\langle n|e^{-Q}|x,y\rangle\langle x,y|m \rangle=\int dx dy \psi^c_n(x,y){\cal C}_n\psi_m(x,y)=\delta(m,n)$.

To conclude, we see that the claim that string theory is the only consistent quantum theory of gravity is not correct. While it might be the only consistent theory which is based on the Einstein equations, once one bases gravity on conformal gravity instead, an entirely different  quantum gravity theory is obtained. Conformal gravity (and by extension conformal supergravity as well) thus allow the construction of a completely consistent quantum theory of gravity in four spacetime dimensions. The author has benefitted immensely from his collaborations with Dr. A. Davidson and Dr. C. M. Bender.

\end{document}